\documentclass[aps,prr,twocolumn,superscriptaddress,showpacs]{revtex4-2}

\usepackage{graphicx}
\usepackage{amsmath,amssymb}
\usepackage{bm}
\usepackage{color}

\makeatletter
\def\btt#1{\texttt{\@backslashchar#1}}%
\DeclareRobustCommand\bblash{\btt{\@backslashchar}}%
\makeatother

\begin{document}

\title{Magnetic excitations in an ionic spin-chain system with a non-magnetic ferroelectric instability}

\author{K. Sunami}
\email{e-mail: sunami@mdf2.t.u-tokyo.ac.jp}
\affiliation{Department of Applied Physics, University of Tokyo, Bunkyo-ku, Tokyo 113-8656, Japan}

\author{Y. Sakai}
\affiliation{Department of Applied Physics, University of Tokyo, Bunkyo-ku, Tokyo 113-8656, Japan}

\author{R. Takehara}
\affiliation{Department of Applied Physics, University of Tokyo, Bunkyo-ku, Tokyo 113-8656, Japan}

\author{H. Adachi}
\affiliation{Department of Applied Physics, University of Tokyo, Bunkyo-ku, Tokyo 113-8656, Japan}

\author{K. Miyagawa}
\affiliation{Department of Applied Physics, University of Tokyo, Bunkyo-ku, Tokyo 113-8656, Japan}

\author{S. Horiuchi}
\affiliation{Research Institute for Advanced Electronics and Photonics (RIAEP), National Institute of Advanced Industrial Science and Technology (AIST), Tsukuba, Ibaraki, 305-8565, Japan}

\author{K. Kanoda}
\email{e-mail: kanoda@ap.t.u-tokyo.ac.jp}
\affiliation{Department of Applied Physics, University of Tokyo, Bunkyo-ku, Tokyo 113-8656, Japan}

\date{\today}

\begin{abstract}
Cross-correlation between magnetism and dielectric is expected to offer novel emergent phenomena. Here, magnetic excitations in the organic donor-acceptor spin-chain system, TTF-BA, with a ferroelectric ground state is investigated by $^1$H-NMR spectroscopy. A nonmagnetic transition with a ferroelectric order is marked by sharp drops in NMR shift and nuclear spin relaxation rate $T_1^{-1}$ at 53 K. Remarkably, the analyses of the NMR shift and $T_1^{-1}$ dictate that the paramagnetic spin susceptibility in TTF-BA is substantially suppressed from that expected for the 1D Heisenberg spins. We propose that the spin-lattice coupling and the ferroelectric instability cooperate to promote precursory polar singlet formation in the ionic spin system with a nonmagnetic ferroelectric instability.
\end{abstract}


\maketitle

\section{Introduction}
One-dimensional (1D) systems possess coupled electron-lattice instabilities, a metal-insulator (Peierls) transition for itinerant electron systems and a paramagnetic-nonmagnetic (spin-Peierls) transition for localized spin systems \cite{Peierls_1955, Cross_1979}. Organic charge-transfer complexes of quasi-1D nature are representative platforms for the study of these issues. Among them, TTF-CA (tetrathiafulvalene-chloranil) composed of 1D mixed stacks of donor molecules, TTF, and acceptor molecules, CA, is a fascinating material showing a neutral-ionic (NI) transition accompanied by a symmetry-breaking lattice dimerization at 81 K under ambient pressure \cite{Torrance_1981, Torrance_1981_1}; it switches from a paraelectric neutral phase (TTF$^{+\rho}$-CA$^{-\rho}$ with $\rho \sim$ 0.3) to a ferroelectric ionic phase ($\rho \sim$ 0.6-0.7). On the other hand, the analogous material, TTF-BA (tetrathiafulvalene-bromanil), in which Cl atoms in CA molecules are substituted by Br atoms [Fig. 1(a)], is in a highly ionic state ($\rho \sim$ 0.95) at all temperatures \cite{Girlando_1985} with every molecule accommodating a $S$ = 1/2 spin, which is paramagnetic at room temperature. Upon cooling, TTF-BA exhibits a non-magnetic dimerization transition with ferroelectricity \cite{Kagawa_2010, Garcia_2005}, putatively a spin-Peierls transition, at 53 K \cite{Girlando_1985}. 

Recent under-pressure studies of TTF-CA have revealed that the ionic phase above $\sim$9 kbar is non-dimerized and paraelectric around room temperature and undergoes a dimerized ferroelectric transition on cooling \cite{Cointe_2017, Takehara_2018}, evoking a view that TTF-BA is equivalent to TTF-CA under pressures above $\sim$9 kbar. Notably, it has recently been shown that the magnetism and conductivity in the paraelectric ionic phase of TTF-CA under pressure are attributed to spin and charge solitons \cite{Sunami_2018, Takehara_2019}. As shown in Fig. 1(b), however, the resistivity in TTF-BA (measured in the present study) is six to seven orders of magnitude larger than in TTF-CA under pressure \cite{Sunami_submitted}, indicating the absence of charge-soliton excitations in TTF-BA; thus, TTF-BA offers a distinct localized spin system with a polarized non-magnetic ground state.

\begin{figure}
\includegraphics[width=8.7cm,clip]{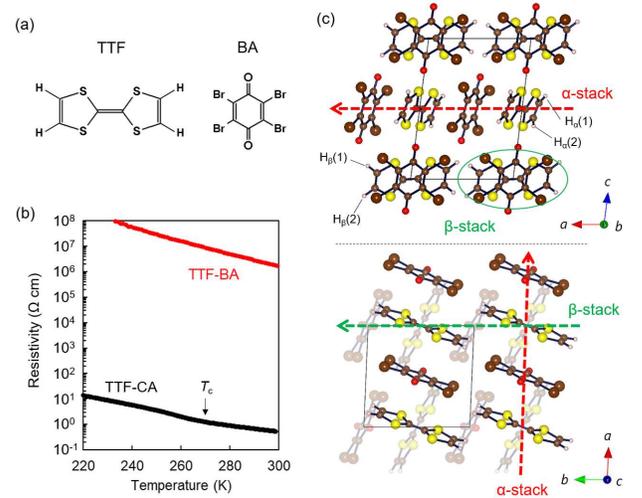}
\caption{(a) Molecular structures of TTF and BA. (b) Temperature dependence of electrical resistivity in TTF-BA under 5 kbar (red line) and TTF-CA under 14 kbar \cite{Sunami_submitted} (black line) measured by the four-terminal method. The resistance in TTF-BA at ambient pressure is too high ($> 10^6$ $\Omega$) to measure by the four-terminal method, and thus we present the 5-kbar data. $T_\mathrm{c}$ is the ferroelectric transition temperature in TTF-CA at 14 kbar. (c) Crystal structures of TTF-BA viewed from the $b$ (upper) and $c$ (lower) axes \cite{Garcia_2005}.
}
\label{Fig1} 
\end{figure}

The present study aims to reveal the nature of spin excitations in TTF-BA, a polarizable Heisenberg antiferromagnetic spin-chain system with a non-magnetic ferroelectric transition, by $^1$H-NMR measurements. In general, $^1$H-NMR has insufficient sensitivity for probing the electronic state because of small hyperfine coupling and molecular motions unwantedly contributing to the NMR relaxation rate. TTF-BA, however, has sizable $^1$H hyperfine coupling constants, no motional molecular parts and, further, appreciably large static and dynamical spin susceptibilities as described later. Owing to all these features of TTF-BA, $^1$H-NMR is competent for probing the spin states.

\begin{figure}
\includegraphics[width=8.7cm, clip]{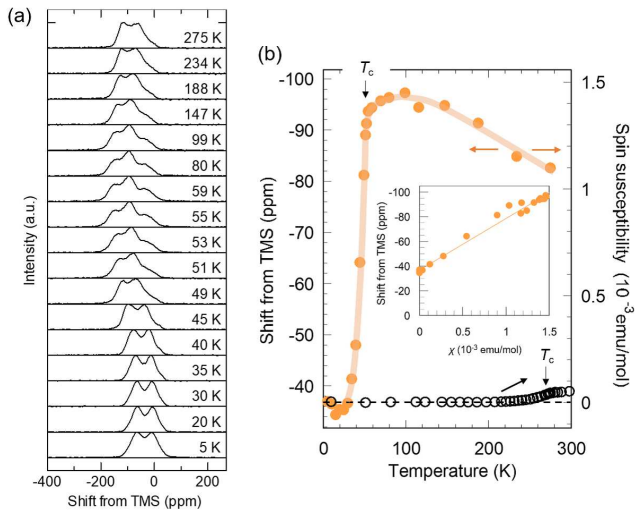}
\caption{Temperature profiles of the $^1$H-NMR spectra (a) and the 1st moments of spectra in TTF-BA (closed yellow circles) (b). In (b), the right axis denotes the scale of the spin susceptibility $\chi$ and the open black circles represent $\chi$ in TTF-CA under 14 kbar \cite{Sunami_submitted}. Inset of (b): Plot of spectral shift vs. spin susceptibility reported in Ref. \cite{Kagawa_2010} in TTF-BA. 
}
\label{Fig2}
\end{figure}

\section{Experimental}
We performed $^1$H-NMR measurements on a single crystal of TTF-BA under the magnetic field $H$ of 3.7 T applied parallel to the $b^*$ axis (perpendicular to the $ac$ plane). In this field configuration, there are four nonequivalent $^1$H nuclei above $T_\mathrm{c}$ [denoted by H$_{\alpha}$(1) and H$_{\alpha}$(2) on the $\alpha$-stack running along the $a$ axis and H$_{\beta}$(1) and H$_{\beta}$(2) on the $\beta$-stack along the $b$ axis in Fig. 1(c)]. For acquiring NMR signals, we employed the so-called solid-echo pulse sequence. The origin of the NMR line shift corresponds to the resonance frequency for TMS (tetramethylsilane). The spin-lattice relaxation rate $T_1^{-1}$ was determined by fitting the stretched exponential function to the relaxation curve of the nuclear magnetization obtained using the standard saturation method. The relaxation curve is nearly single-exponential except below the non-magnetic transition temperature, where a somewhat non-single exponential feature appears very probably due to the orphan spins failing to form singlets and/or minor impurity spins as seen later. The frequency dependence of $T_1^{-1}$ was measured in the range of 20-370 MHz which corresponds to 0.5-8.6 T.

\section{Results}
\subsection{NMR spectra}
$^1$H-NMR spectrum at around room temperature is formed by slightly asymmetric two broad peaks [Fig. 2(a)]. Upon cooling, the spectrum changes its shape but turns into symmetric two peaks below $\sim$50 K. The spectral shift defined by the first moment of the entire spectrum is plotted in Fig. 2(b); it shows a maximum around $\sim$100 K and a sharp decrease indicating the non-magnetic transition at $T_\mathrm{c}$ = 53 K. The spectral shift is contributed by the spin shift, proportional to the spin susceptibility $\chi$, and the temperature-independent chemical shift. These are separated by plotting the shift values against the previously reported $\chi$ values \cite{Kagawa_2010} [inset of Fig. 2(b)]; the slope of the linearity gives the hyperfine coupling component parallel to the field direction averaged for the four $^1$H sites, $a_{\parallel}^{\mathrm{ave}}$, of $-0.47$ kOe/$\mu_\mathrm{B}$ and the intercept determines the chemical shift of $-37$ ppm, which has uncertainty of tens ppm but is not involved in the present analysis. The hyperfine coupling tensors have anisotropies arising from the dipolar interactions between nuclear and electron spins. We evaluate a dipolar field at each $^1$H site generated by the electron spin as follows; (i) a magnetic moment of 1 $\mu_\mathrm{B}$ is distributed over the atomic sites (precisely, nuclear positions) in a molecule according to the Mulliken populations calculated by the extended H$\mathrm{\ddot{u}}$ckel method \cite{Mori_1984} using the atomic coordinates at 293 K reported in Ref. \cite{Garcia_2005} and (ii) we incorporate dipole fields from electron spins on six neighboring molecules. For the present field configuration ($H \parallel b^*$), the dipole hyperfine coupling components parallel to the field direction are calculated to be 0.35, 0.41, $-0.22$, $-0.32$ kOe/$\mu_\mathrm{B}$ for H$_{\alpha}$(1), H$_{\alpha}$(2), H$_{\beta}$(1) and H$_{\beta}$(2), respectively, whose average $(a_{\parallel}^{\mathrm{aniso}})^{\mathrm{ave}}$ is 0.05 kOe/$\mu_\mathrm{B}$. The observed value of $-0.47$ kOe/$\mu_\mathrm{B}$ is contributed by the dipole (anisotropic) part, $(a_{\parallel}^{\mathrm{aniso}})^{\mathrm{ave}}$, and an isotropic part, $a^{\mathrm{iso}}$, which thus yields $-0.52$ $(= -0.47-0.05)$ kOe/$\mu_\mathrm{B}$. The asymmetric shape of spectra above $T_\mathrm{c}$ possibly arise from the different anisotropic local fields at the four $^1$H sites as estimated above. Below $T_\mathrm{c}$, where the spin shift vanishes, the spectra take a shape of the so-called Pake doublet derived from the nuclear dipolar interactions between H$_{\alpha(\beta)}$(1) and H$_{\alpha(\beta)}$(2), possibly broadened by the anisotropic chemical shift tensor giving site-dependent shifts.

\subsection{Spin-lattice relaxation rate $T_1^{-1}$}
The non-magnetic transition was also captured by a steep decrease in the spin-lattice relaxation rate $T_1^{-1}$ below $T_\mathrm{c}$ [Fig. 3(a)]. The drop of $T_1^{-1}$ just below $T_\mathrm{c}$ is characterized by an activation energy $\Delta_{T_1^{-1}}$ of 240 K, which is comparable with the activation value $\Delta_s$, 300 K, characterizing the drop in the spin shift multiplied by $T$ just below $T_\mathrm{c}$, as shown in Fig. 3(b). A levelling-off in $T_1^{-1}$ below 35 K is likely caused by free spins failing to form singlets and/or minor impurity spins, as often observed in $^1$H-NMR for spin-singlet phases \cite{Itou_2009}; the exponent in the stretched exponential fitting of the nuclear relaxation curve, which is nearly unity above 35 K, decreases to 0.8 at 15K [inset of Fig. 3(a)], also suggesting inhomogeneous nuclear relaxations by the dispersed orphan spins. The near agreement between the $\Delta_{T_1^{-1}}$ and $\Delta_s$ values is consistent with the singlet-triplet excitations unlike in the ferroelectric phase of TTF-CA showing a clear disagreement, which indicates that polaron excitations inheriting charge solitons vitally excited above $T_\mathrm{c}$ and singlet-triplet excitations occur in different energy scales and contribute to shift and $T_1^{-1}$ with distinct weights \cite{Sunami_submitted}. The contrasting behaviors of TTF-CA and TTF-BA is reasonable because TTF-CA is situated near the NI phase boundary with the charge-transfer instability whereas TTF-BA is an ionic Mott insulator with a large charge gap of $\sim$8000 K \cite{Tokura_1989}, thus carrying low energy excitations only in the spin degrees of freedom.

\begin{figure}
\includegraphics[width=8.9cm]{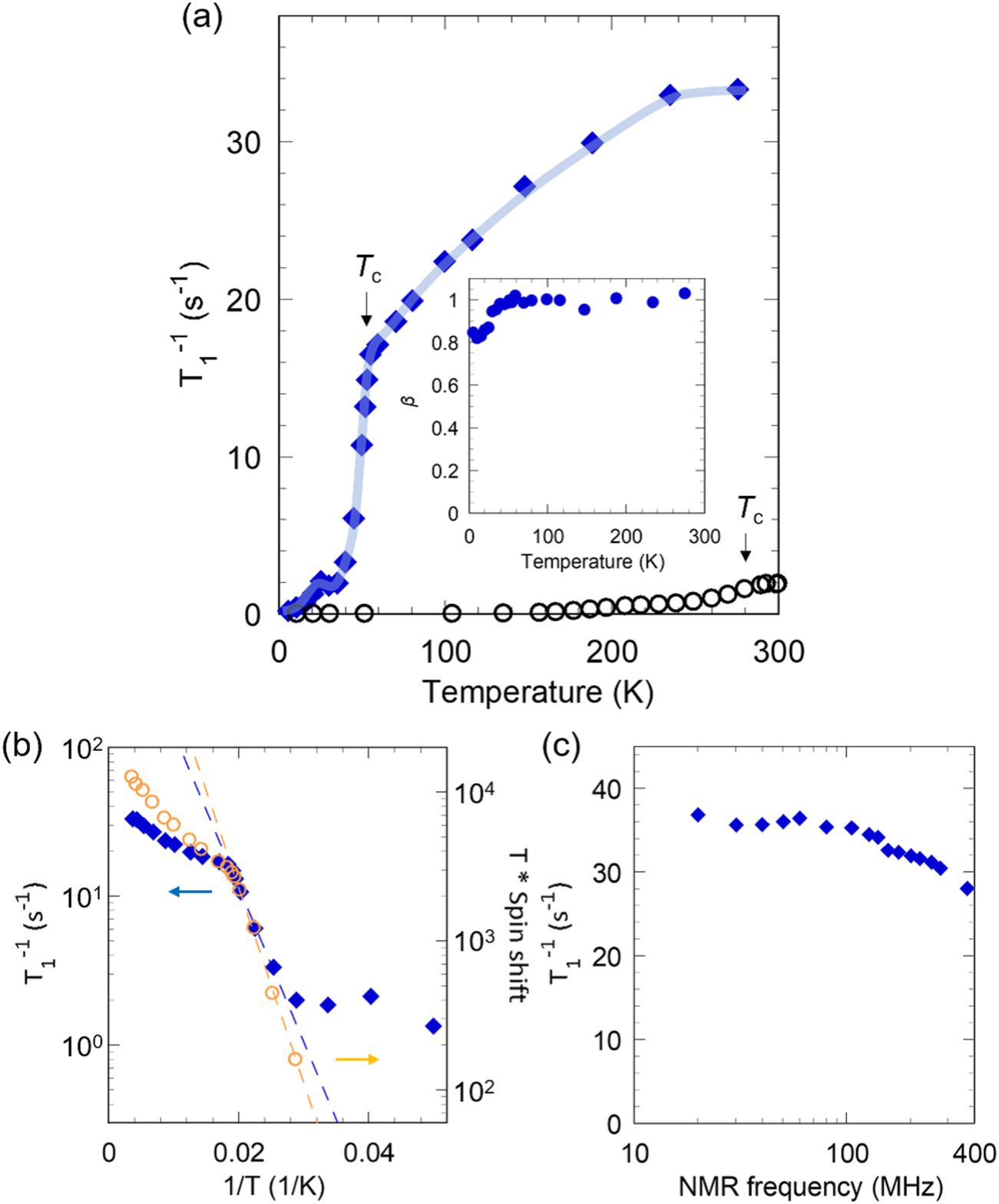}
\caption{(a) Temperature dependence of the $^1$H-NMR spin-lattice relaxation rate $T_1^{-1}$ in TTF-BA (closed blue diamonds) and TTF-CA under 13 kbar (open black circles) reported in Ref. \cite{Sunami_2018}. Inset: Temperature dependence of the exponent $\beta$ in the stretched exponential fitting of the nuclear relaxation curves for TTF-BA. (b) Activation plots of $T_1^{-1}$ (closed blue diamonds) and spin shift multiplied by $T$ (open yellow circles) in TTF-BA. The broken lines indicate single exponential fits to the data in 35 K $< T < T_\mathrm{c}$. (c) Frequency dependence of $T_1^{-1}$ at 280 K in TTF-BA.
}
\label{Fig3} 
\end{figure}

In the paramagnetic phase above $T_\mathrm{c}$, the spin susceptibility and $T_1^{-1}$ are much larger than those of TTF-CA under pressure [Figs. 2(b) and 3(a)], implying that TTF-BA is regarded as a $S$ = 1/2 localized spin system, not a soliton matter as in the ionic paraelectric phase of TTF-CA \cite{Sunami_2018}. This picture is supported by the frequency-insensitive $T_1^{-1}$ observed at 280 K [Fig. 3(c)], which is compared to the pronounced frequency dependence of $T_1^{-1}$ pointing to the diffusive motion of spin solitons \cite{Sunami_2018}. We note that, in 1D Heisenberg spin systems for high-temperature limit, $T_1^{-1}$ is also expected to show the frequency dependence due to the classical spin diffusion \cite{Kikuchi_2001, Smith_1976}; thus, $T_1^{-1}$ insensitive to frequency implies that TTF-BA cannot be regarded as the pure 1D spin system, which is discussed in more detail below, although the slight decrease of $T_1^{-1}$ at higher frequencies is possibly derived from the weak 1D nature of spin chain.

\subsection{Evaluations of exchange interactions}
The lattice structure of TTF-BA is rather complicated as seen in Fig. 1(c), where the 1D chains running along the $a$ and $b$ axes stack alternatively along the $c$ direction with short TTF-TTF inter-column contacts, quite different from TTF-CA where the 1D chains are arranged in parallel to each other. We evaluated intermolecular transfer integrals in TTF-BA through molecular orbital calculations \cite{Mori_1984}. The largest transfer integral is $t_{\parallel}$ = 43-46 meV between the TTF-HOMO and BA-LUMO along the $a$ or $b$ axis [the bold lines in Fig. 4(a)], the next largest one is $t_{\perp}$ = 28 meV between the TTF-HOMOs along the $c$ direction [the double lines in Fig. 4(a)], and the third largest one is $t_{\perp}'$ = 19 meV between the TTF-HOMO in the $\alpha (\beta$) chain and the CA-LUMO in the $\beta (\alpha)$ chain [the single lines in Fig. 4(a)]. Other values, for instance the transfer integrals between the adjacent $\alpha (\beta$) columns in parallel [the dotted lines in Fig. 4(a)], are below $\sim$10 meV. For modelling the spin network of TTF-BA, we calculated the exchange interaction $J$ between the localized spins using the two expressions derived within the second-order perturbation in the ionic Hubbard model \cite{Katsura_2009}; $J = 2t^2/(U-\Delta)+2t^2/(U+\Delta)$ for a donor-acceptor pair and $J = 4t^2/U$ for the donor-donor (acceptor-acceptor) pair, where $\Delta$ is the effective energy difference between the TTF-HOMO and BA-LUMO levels including the inter-site Coulomb energy and $U$ is the on-site Coulomb energy. In this model, the charge transfer gap is expressed by $U - \Delta$, which is 0.8 eV according to the infrared \cite{Girlando_1985} and resistivity \cite{Tokura_1989} measurements. With $U$ = 1.5 eV employed in the theoretical calculations \cite{Nagaosa_1986}, we obtained $\Delta$ = 0.7 eV, which yields $J_{\parallel} \equiv J_{\mathrm{TTF}(\alpha)-\mathrm{BA}(\alpha)}$ (or $J_{\mathrm{TTF}(\beta)-\mathrm{BA}(\beta)}$) = 73-84 K, $J_{\perp} \equiv J_{\mathrm{TTF}(\alpha)-\mathrm{TTF}(\beta)}$ = 24 K, and $J_{\perp}' \equiv J_{\mathrm{TTF}(\alpha)-\mathrm{BA}(\beta)}$ (or $J_{\mathrm{TTF}(\beta)-\mathrm{BA}(\alpha)}$) = 14 K [Fig. 4(a)]. Bewick $et$ $al$. computed $t_{\parallel}$ = 60 meV based on the semi-empirical theory \cite{Bewick_2006}, which leads to $J_{\parallel}$ = 140 K, not far from the above calculation.

On the other hand, the exchange interactions can be evaluated from the experimental data of $T_1^{-1}$ using the following formula for localized spins in the high-temperature limit \cite{Moriya_1956},
\begin{equation}
T_1^{-1} = \sqrt{\frac{\pi S(S+1)}{3Z}} \frac{g^2 \hbar \gamma_\mathrm{N}^2 a_\perp^2}{J},
\end{equation}
where $Z$ is the number of the neighboring sites, $g$ is the electron $g$-factor, $\gamma_\mathrm{N}$ is the nuclear gyromagnetic ratio, $\hbar$ is the reduced Planck constant, and $a_\perp$ is the hyperfine coupling component perpendicular to the field direction given by
\begin{equation}
\begin{split}
a_\perp^2 = \frac{1}{2} \left[(a_{yy}^2+a_{zz}^2 ) H_x^2+(a_{zz}^2+a_{xx}^2 ) H_y^2+(a_{xx}^2+a_{yy}^2)H_z^2\right],
\end{split}
\end{equation}
where $a_{ii}$ ($i$ = $x$, $y$, and $z$) is the principal value of the hyperfine coupling tensor and $H_i$ is the direction cosine of the applied field for respective $^1$H sites. In calculating $a_\perp$, we employed the $a^{\mathrm{iso}}$ value of $-0.52$ kOe/$\mu_\mathrm{B}$ as the isotropic part and calculated the anisotropic dipole part from the on-molecular spin, which yields ($-0.3$$a^{\mathrm{aniso}}$, $-1.7$$a^{\mathrm{aniso}}$, 2$a^{\mathrm{aniso}}$) with $a^{\mathrm{aniso}}$ $\sim$ 0.27 kOe/$\mu_\mathrm{B}$ in principal values. The principal axes of the hyperfine tensor differ among the four $^1$H sites so that $H_i$ depends on the $^1$H site. The average of $a_\perp^2$ over the four $^1$H sites yields ($a_\perp^2)^\mathrm{ave}$ $\sim$ 0.44 kOe$^2$/$\mu_\mathrm{B}^2$ for the present field configuration of $H \parallel b^*$. Considering the inter-chain exchange interaction, we replace $J$ to $\sqrt{J_\parallel^2+J_\perp^2+J_\perp'^2}$ in Eq. (1) with $J_{\parallel}$ = 3-6$J_{\perp}$ = 5-10$J_{\perp}'$ and $Z$ = 2. The high-temperature limit of $T_1^{-1}$ is estimated at 41 s$^{-1}$ by extrapolating it to 1/$T$ = 0 in Fig. 3(b). Substituting these values to Eq. (1), we obtained $J_{\parallel}$ $\sim$ 140 K, which is nearly in agreement with $J_{\parallel}$ = 73-140 K calculated from the second-order perturbation formulae.

\begin{figure}
\includegraphics[width=7.5cm]{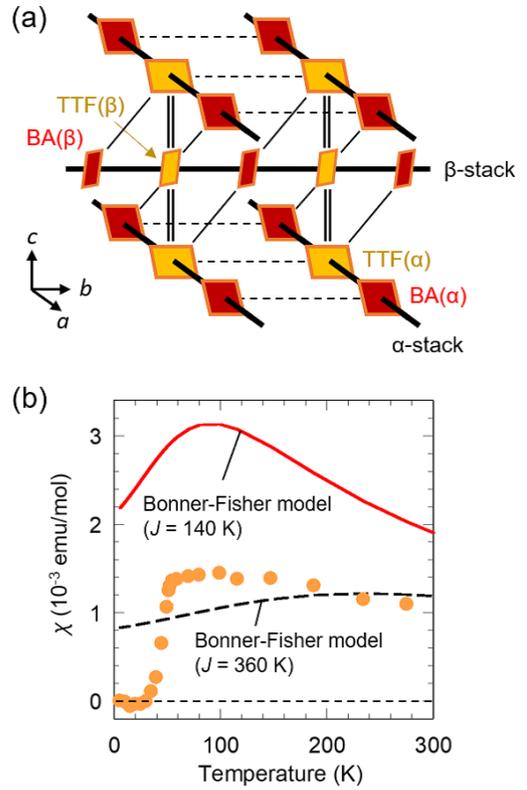}
\caption{(a) Network of transfer integrals (or the exchange interactions) in TTF-BA above $T_\mathrm{c}$. The bold, double, and single lines represent the nearest ($t_{\parallel}$, $J_{\parallel}$), second-nearest ($t_{\perp}$, $J_{\perp}$), and third-nearest ($t_{\perp}'$, $J_{\perp}'$) neighboring molecular pairs, respectively. (b) Comparison between the experimental spin susceptibility and the Bonner-Fisher model calculations for $J$ = 140 K (experimentally determined) and $J$ = 360 K (for reference).
}
\label{Fig4}
\end{figure}

\section{Discussion}
Anomalous spin excitations in TTF-BA are highlighted by comparing the spin susceptibilities derived from the spin shifts with the Bonner-Fisher curve expected in 1D antiferromagnetic (AF) Heisenberg spin systems \cite{Bonner_1964, Estes_1978} with $J$ = 140 K obtained above. Bewick $et$ $al$. calculated the spin susceptibility in the 1D mixed-stack charge-transfer systems described by the Peierls-Hubbard model \cite{Bewick_2006_1}, the temperature profile of which is nearly close to the Bonner-Fisher curve for the ionic limit ($\rho \to 1$). However, on the basis of the experimentally determined $J$ value, a large difference between the experimental data and the Bonner-Fisher curve is evident in TTF-BA with $\rho$ $\sim$ 0.95. We note that, for any $J$ value, the Bonner-Fisher curve does not reproduce the experimental susceptibility as reported in Refs. \cite{Girlando_1985, Kagawa_2010} and shown in Fig. 4(b). In general, the spin susceptibility is more suppressed as the number of the neighboring sites increase, and thus the complicated spin network in TTF-BA appears responsible for the reduction of the susceptibility from that of the pure 1D spin chain. However, around temperatures of $\sim$$J_{\parallel}$ and below, the intra-chain AF correlations are developed so that, on a site in adjacent chains, the exchange fields produced through $J_{\perp}$ and $J_{\perp}'$ [see Fig. 4(a)] are offset by each other due to the frustration between $J_{\perp}$ and $J_{\perp}'$. Thus, the spin network is expected to become 1D-like below $\sim$140 K; then, the remarkable discrepancy between the experimental and calculated behaviors in TTF-BA invokes unusual mechanism of spin excitations.

We discuss the origins of the discrepancy in the light of the distinctive nature of TTF-BA as a localized spin system. The first one is the enhanced spin-lattice coupling. Theoretically, working with the semiclassical treatment of bosonized Hamiltonian for NI transition systems, Tsuchiizu $et$ $al$. suggested that the spin susceptibility in the 1D AF Heisenberg spin systems with spin-lattice couplings and site-alternating potentials does not follow the Bonner-Fisher curve but carry unconventional excitations, called ``spin polarons'', unlike the conventional spinon excitations \cite{Tsuchiizu_2016}. In particular, the spin-lattice coupling is expected to cause the local spin-singlet pairings even above $T_\mathrm{c}$ through the lattice fluctuations. Another exclusive feature of the present spin system is electric polarizability leading to polar fluctuations inherent in the mixed-stack ionic Mott insulator. Kagawa $et$ $al$. reported that the conventional relationships between $\Delta$ vs. $T_\mathrm{c}$ and between $H$ vs. $T_\mathrm{c}$ in spin-Peierls systems are broken in TTF-BA \cite{Kagawa_2010}; typically, $\Delta/k_\mathrm{B}T_\mathrm{c}$ = 2.5 \cite{Orignac_2004} and $\alpha$ = 0.38 \cite{Cross_1979_1} in the spin-Peierls systems, whereas $\Delta/k_\mathrm{B}T_\mathrm{c}$ = 4.3 and $\alpha$ = 0.18 in TTF-BA, where $\alpha$ is the coefficient given by $1- T_\mathrm{c}(H)/T_\mathrm{c}(0) = \alpha[g\mu_\mathrm{B}H/2k_\mathrm{B}T_\mathrm{c}(0)]^2$. The large $\Delta/k_\mathrm{B}T_\mathrm{c}$ value suggesting the spin gap far exceeding the energy scale of the transition and the small $\alpha$ value signifying outstanding robustness to magnetic field invoke an additional mechanism to stabilize the non-magnetic phase beyond the conventional spin-Peierls framework. Polar fluctuations, which cause local donor acceptor pairing, possibly favor precursory singlet formation in the paramagnetic phase, suppressing spin susceptibility above $T_\mathrm{c}$.

It is a highly likely view that the precursory singlet fluctuations observed in TTF-BA manifest themselves in an extreme manner as a dimer liquid in TTF-CA, where most of the TTF and CA molecules form polar singlet pairs, whose long-range order is interrupted by soliton excitations. TTF-BA with a charge transfer of $\rho$ $\sim$ 0.95 is a nearly perfect ionic ferroelectric, whereas TTF-CA with $\rho$ $\sim$ 0.6-0.7 is an intermediately charge-transferred electronic ferroelectric. It is intriguing to see how the donor-acceptor spin chain system vary its magnetic excitations from the local spin regime to the soliton regime when $\rho$ is reduced from unity.

\section{Conclusion}
In conclusion, $^1$H-NMR spectroscopy of the alternating donor-acceptor ionic chain system, TTF-BA, revealed that it hosts an extraordinary spin system, neither understandable by the soliton matter realized in the analogous system TTF-CA nor the conventional Heisenberg spin system with the spin-Peierls instability. The spin shift and nuclear spin-lattice relaxation rate clearly captured the spin-singlet transition at 53 K. The paramagnetic state above 53 K is demonstrated to host localized spin chains. However, the analyses of the spin shift and $T_1^{-1}$ found a substantial reduction in the spin susceptibility from that expected for the antiferromagnetic Heisenberg spin chains, evoking a view that TTF-BA offers an exclusive ionic spin system with unusually coupled magnetic and polar fluctuations enhanced prior to a non-magnetic ferroelectric order. This potentially novel cross-correlated fluctuation, which is possibly a generic feature for extensive ionic spin systems, is a profound addition to the correlated electron physics, awaiting theoretical challenges to treat jointly spin dynamics and electric polar correlation.

\acknowledgments
This work was supported by the JSPS Grant-in-Aids for Scientific Research (grant nos. JP17K05532, JP18H05225 and JP19H01846) and by the Murata Science Foundation.



\end{document}